\documentclass{article}

\usepackage{arxiv}

\usepackage[utf8]{inputenc} % allow utf-8 input
\usepackage[T1]{fontenc}    % use 8-bit T1 fonts
\usepackage{hyperref}       % hyperlinks
\usepackage{url}            % simple URL typesetting
\usepackage{booktabs}       % professional-quality tables
\usepackage{amsfonts}       % blackboard math symbols
\usepackage{nicefrac}       % compact symbols for 1/2, etc.
\usepackage{microtype}      % microtypography
\usepackage{lipsum}
\usepackage{comment}
\usepackage{tikz}
\usepackage{float}
\usepackage{amsmath}
\usetikzlibrary{arrows.meta}
\tikzset{%
  >={Latex[width=2mm,length=2mm]},
            base/.style = {rectangle, rounded corners, draw=black,
                           minimum width=4cm, minimum height=1cm,
                           text centered, font=\sffamily},
  activityStarts/.style = {base,minimum width=2.5cm, fill=blue!30},
       startstop/.style = {base,minimum width=2.5cm, fill=red!30},
    activityRuns/.style = {base,minimum width=2.5cm, fill=green!30},
         process/.style = {base, minimum width=2.5cm, fill=orange!15,
                           font=\ttfamily},
}

\title{Mathematical Models for Describing and Predicting the COVID-19 Pandemic Crisis}

\author{
  Cintra, P. H. P.\thanks{Use footnote for providing further
    information about author (webpage, alternative
    address)---\emph{not} for acknowledging funding agencies.} \\
  Instituto de Física\\
  Universidade de Brasilia\\
  Brasilia, DF, Brasil, 70910-900 \\
  \texttt{pedrohpc96@hotmail.com} \\
   \And
 Citeli, M. F. \\
  Instituto de Física\\
  Universidade de Brasilia\\
  Brasilia, DF, Brasil, 70910-900 \\
  \texttt{miguelciteli@gmail.com} \\
  \And
  Fontinele, F. N. \\
   Instituto de Física\\
   Universidade de Brasilia\\
   Brasilia, DF, Brasil, 70910-900\\
   \texttt{feradofogo@hotmail.com}
}

\begin{document}
\maketitle

\begin{abstract}
The present article studies the extension of two deterministic models for describing the novel coronavirus pandemic crisis, the SIR model and the SEIR model. The models were studied and compared to real data in order to support the validity of each description and extract important information regarding the pandemic, such as the basic reproductive number $\mathcal{R}_0$, which might provide useful information concerning the rate of increase of the pandemic predicted by each model. We next proceed to making predictions and comparing more complex models derived from the SEIR model with the SIRD model, in order to find the most suitable one for describing and predicting the pandemic crisis. Aiming to answer the question if the simple SIRD model is able to make reliable predictions and deliver suitable information compared to more complex models.
\end{abstract}

% keywords can be removed

\section{Introduction}
Back on 2015, a group of researches described the potential for a SARS coronavirus circulating inside bats to mutate to humans \cite{menachery2015sars}. Early on 2020 the world suffers from an new pandemic crisis caused by the novel coronavirus, SARS-CoV-2, belonging to the \textit{Betacoronavirus} genus and with probable origin on bats \cite{lu2020genomic}. The first cases of the novel virus date back to December 2019 at the Food Market of Wuhan, China \cite{li2020early}, where bats are sold among other exotic animals, since then the virus has been spreading throughout China, later Asia, Europe, Africa and America, causing a global scale economic crisis and being notified by the World Health Organization as an pademic on March 11th.

COVID-19 is a respiratory disease caused by the Sars-Cov-2 virus, currently in human-to-human sustained transmission \cite{world2020coronavirus}. We know that the contamination mainly occurs by a close interaction with infected individuals, as the viral charge is carried by respiratory droplets that can remain in suspension in air or deposited on surfaces of common contact. As a novel strain of the \textit{Coronaviridae} family, it is not expected that any individual has antibodies against it, which causes the entire population to be susceptible to infection. As an individual is exposed to the virus, the incubation period begins, with no symptoms and a small chance of contaminating others. When the virus is onset, the infected individual show symptoms in a varied range of intensities and may develop severe acute respiratory syndrome. COVID-19 has a general mortality rate bellow 5\% \cite{cascella2020features}, with an average of 2.3\%. The behavior of the disease is age dependent, with the higher risk group being older populations, that present a mortality rate of 8\% for individuals between 70-79 years and 14.8\% for people older than 80 years \cite{surveillances2020epidemiological}. However, even with a low mortality rate, the number of hospitalizations is quite high, with 5\% of the cases being critical and 14\% being severe \cite{wu2020characteristics}, presenting an challenge to health care systems of some countries.

Mathematical models predicted the potential for an international outbreak early on \cite{wu2020nowcasting} and described how Wuhan became the center of an epidemic crisis on China. The outbreak quickly spread throughout mainland China and other countries. Although the pandemic crisis began on Wuhan, today the United States of America is the epicenter of the pandemic crisis on the world.

Many models are widely used from scenario prediction \cite{fernandez2020estimating, iwata2020simulation, shoukat2020projecting, walker2020global} to data analysis \cite{dehning2020inferring}. Among all those models, the most used ones are the SIR and SEIR models \cite{anastassopoulou2020data}, including their modifications to include hospitalizations, asymptomatic cases and other compartments \cite{choi2020estimating}. We develop here a comparative study between modified SIR and SEIR models; in order to provide support for the accuracy of both models, we compare the differences on the fitting process between the simple SIRD and simple SEIRD models and the accuracy of prediction generated by the simple SIRD and the SEIRD model with age division, using data from Germany and the Republic of Korea. The choice was based on the accuracy of the data for representing the true scale and dynamics of the pandemic, other countries who presented a much lower testing rate such as Brazil \cite{cintra2020estimative} are not reliable sources for testing models describing the disease. There are also countries such as Taiwan and Iceland, which kept track of the disease; however the number of cases in those countries were much lower, escaping the deterministic nature of the models described here.

\section{Theory}

Mathematical models for disease epidemic are either deterministic or stochastic \cite{ferrante2016stochastic}, where the first may be considered some sort of thermodynamic limit of the second. An analogy made with thermodynamics, where given a big enough number of particles in a gas, for example, you find deterministic equations to describe the behavior of the gas given by the laws of thermodynamics without the need to know the exact behavior of each particle. Otherwise, when your number of molecules is low or you try to compute too many interactions between particles of your system, the random behavior and fluctuations start taking place and you get to a stochastic model.

We describe here a simple extensions of models constantly used on literature \cite{nesteruk2020estimations, nesteruk2020statistics, iwata2020simulation} and with it show some possible behaviors of a disease outbreak.

\subsection{SIRD}

A simple mathematical model for disease epidemic can been built dividing the population in 3 groups: susceptible individuals (S), infected individuals (I) and recovered individuals (R). The model composed by these groups is called the \textit{SIR model}. In this article, however, we consider also individuals who have died by the disease, denote by D. Following the same arguments of the \textit{SIR} model, the \textit{SIRD} model can be described by the set of four differential equations: 
\begin{align}
\label{sird S}
&\frac{dS}{dt} = -\frac{\beta}{N} I(t)S(t) \\
\label{sird I}
&\frac{dI}{dt} = \frac{\beta}{N} I(t) S(t) - (\gamma + \mu) I(t) \\
\label{sird R}
&\frac{dR}{dt} = \gamma I(t) \\
\label{sird D}
&\frac{dD}{dt} = \mu I(t) 
\end{align}

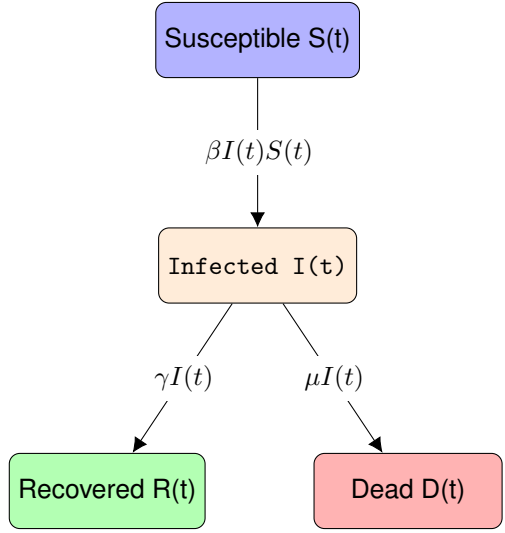
\begin{figure}[H]
    \centering
    \begin{tikzpicture}[node distance=3cm,
    every node/.style={fill=white, font=\sffamily}, align=center]
  % Specification of nodes (position, etc.)
  \node (start)             [activityStarts]              {Susceptible S(t)};
  \node (onCreateBlock)     [process, below of=start]          {Infected I(t)};
  \node (activityRuns)      [activityRuns, below of=onCreateBlock, xshift=-2cm]
                                                      {Recovered R(t)};
    \node (ActivityEnds)      [startstop, below of=onCreateBlock, xshift=2cm]
                                                        {Dead D(t)};
                                    
  % Specification of lines between nodes specified above
  % with aditional nodes for description 
  \draw[->]             (start) -- node[text width=3cm]
                    {$\beta I(t)S(t)$} (onCreateBlock);
  \draw[->]     (onCreateBlock) --
  node[text width=1cm]
                    {$\gamma I(t)$}(activityRuns);
  \draw[->]     (onCreateBlock) -- node[text width=1cm]
                    {$\mu I(t)$}
  (ActivityEnds);
                                   
    \end{tikzpicture}
    \caption{Representation of a SIRD model, a susceptible person gets infected and either dies or recovers from the disease.}
    \label{fig.RepresentationSIRD}
\end{figure}

Last equation is easily understood by thinking that the variation of the number of deaths may be proportional to the infected individuals, where the proportionality constant is denoted by $\mu$. The constants $\gamma$ and $\beta$ are, respectively, the recovery rate and the number of infected, where $\mu$ and $\gamma$ are given in terms of the infection fatality rate (IFR) or the case fatality rate (CFR); that is, the number of people who contracted the disease and died according to the total number of infections (IFR) or the registered number of infections (CFR), and the average time taken from symptoms onset to recovery, $\tau_r$, or death $\tau_d$, formally $\mu = P_{CFR}/\tau_d$ and $\gamma = (1-P_{CFR})/\tau_r$. The equations are simply a mathematical way to describe how individuals passes from one group to the other according to the following chain of events: A susceptible individual becomes infected by the virus, and from this point, it either dies or recovers (Figure \ref{fig.RepresentationSIRD}).

Summing the four equations we get 
\begin{align}
    S(t) + I(t) + R(t) + D(t) = \text{const} ,
\end{align}
where the constant may represent the total number of individuals, $N$. Before proceeding, we propose the initial condition that, when $t$ goes to zero, $I(t) = I_0$, $R(t) = D(t) = 0$, and, therefore, $S(t) = S_0 = N - I_0 \approx N$. Such an assumption is based on the fact that the entire population is susceptible to the SARS-CoV-2 virus.

Since $R(t)$ and $D(t)$ are both data updated day by day in Germany and Korea, it would be helpful to write $I(t)$ as function of them so as to predict its behavior, obtaining, for example, the maximum number of infected individuals. For reasons that may be clear soon, we first write $I(t)$ in terms of $S(t)$. An intuitive step is to divide equation \eqref{sird I} by \eqref{sird S}. Thus, 
\begin{align}
    &\frac{dI/dt}{dS/dt} = -1 + k \frac{1}{S},
\end{align}
where $k = (\gamma + \mu)/\beta$. Eliminating the temporal dependence, we get a separable differential equation, that is, 
\begin{align}
    dI = -dS + k \frac{dS}{S},
\end{align}
which the solution is easily verified to be 
\begin{align}
\label{essa}
    I(t) = -S(t) + k \ln{S(t)} + \text{const}.
\end{align}

Applying the initial condition, we obtain
\begin{align}
    &I_0 = -(N - I_0) + k \ln(N - I_0) + \text{const} \\
    \nonumber
    & \rightarrow \text{const} = N - k\ln(N - I_0).
\end{align}

Hence, equation \eqref{essa} may be written as

\begin{equation}
\label{eq i}
    I(t) = N - S(t) + k \ln \frac{S(t)}{N - I_0}.
\end{equation}

\begin{figure}[H]
    \centering
    \includegraphics{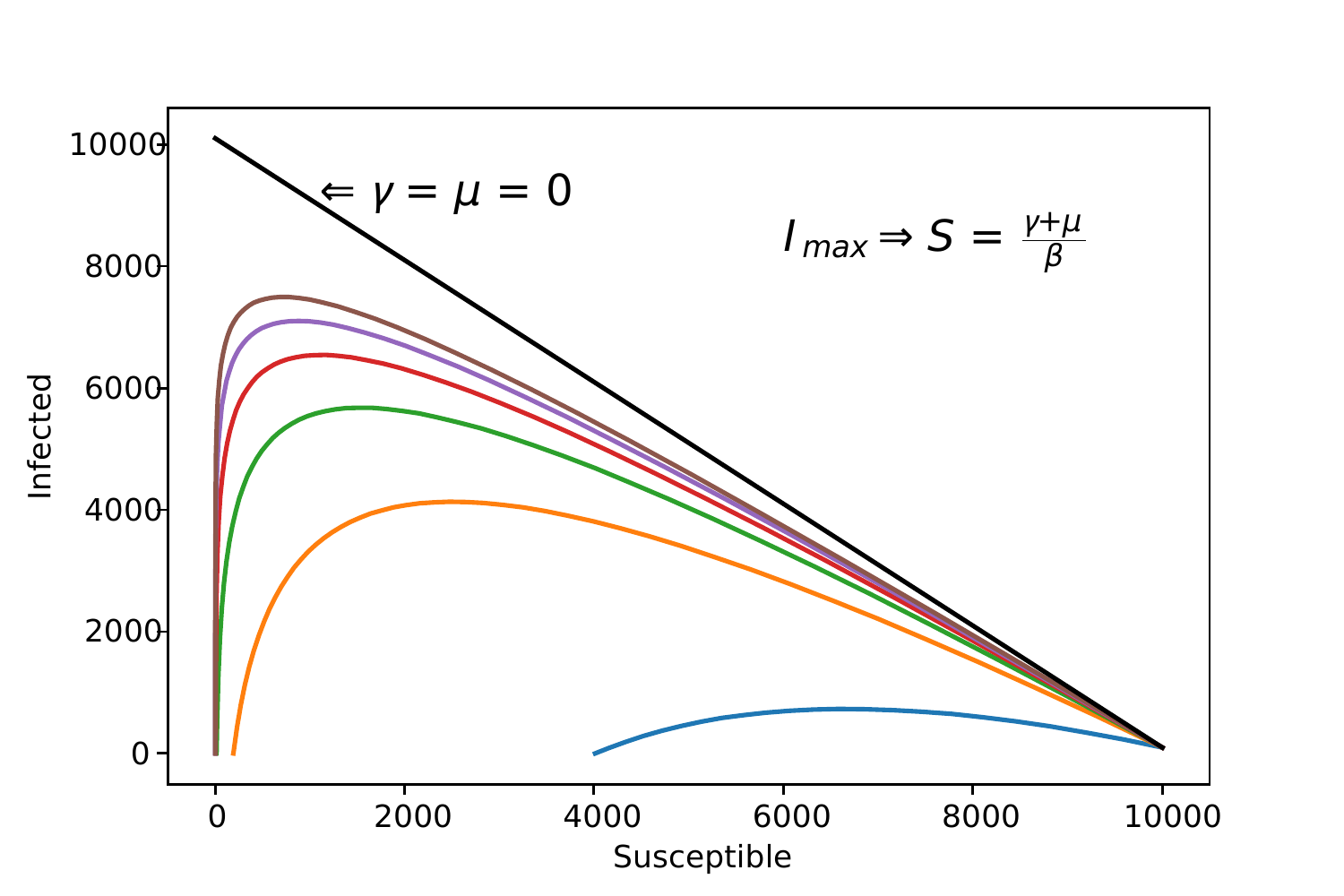}
    \caption{Plot of equation \eqref{eq i} with different conbinations of $\gamma$ and $\mu$.}
    \label{fig.I(S)}
\end{figure}

We can visualize here, that depending on the combination of $\gamma$ and $\mu$, $I$ reaches 0 before the entire population $S$ becomes infected (Figure \ref{fig.I(S)}).

Next, we may write $S(t)$ in terms of $R(t)$ and $D(t)$. For this purpose, we begin by dividing equation (\ref{sird S}) by (\ref{sird R}) and (\ref{sird S}) by (\ref{sird D}), 
\begin{align}
    &\frac{dS}{dR} = -\frac{\beta}{\gamma} S(t) \,\,\, \text{and} \\
    &\frac{dS}{dD} = - \frac{\beta}{\mu} S(t).
\end{align}

Adding these two equations and writing $S(R, D)$ as $S(R, D) = f(R) g(D)$, we get 
\begin{align}
\label{bla}
   \frac{1}{f} \frac{df}{dR} + \frac{1}{g} \frac{dg}{dD} = -\beta \left(\frac{1}{\gamma} + \frac{1}{\mu}\right)
\end{align}

Since \eqref{bla} is a separable equation, the well-known solution is given by 
\begin{align}
\label{solution1}
    &f(R) = A e^{a R} \,\,\, \text{and} \\
\label{solution2}
    &g(D) = B e^{b D}.
\end{align}

Therefore, $S(t)$ can be written as 
\begin{align}
\label{s com a e b}
    S(t) = C e^{aR(t) \, + \,  bD(t)},
\end{align}
where we absorbed both constants $A$ and $B$ into $C$. By the initial condition, we find that $C = N - I_0$. To find $a$ and $b$, we must derive \eqref{s com a e b} in time under the condition that it may return to equation \eqref{sird S}. In this way, we see that
\begin{align}
    a \gamma + b \mu = -\beta. 
\end{align}

By the other hand, substituting equations \eqref{solution1} and \eqref{solution2} in \eqref{bla}, we get 
\begin{align}
    a + b = -\beta \left(\frac{1}{\gamma} + \frac{1}{\mu}\right).
\end{align}

Solving this system, 
\begin{align}
    &a = - \frac{\beta}{\gamma(1 - \gamma/\mu)} \\
    &b = - \frac{\beta}{\mu(1 - \mu/\gamma)}
\end{align}

Hence, $I(t)$ can be finally written as 
\begin{align}
\label{I(t)}
    I(t) = N - (N - I_0) e^{aR(t) + bD(t)} + \gamma \mu \frac{\gamma + \mu}{\gamma - \mu} \left(\frac{R(t)}{\gamma^2} - \frac{M(t)}{\mu^2}\right)
\end{align}

With this equation, see that as $t \rightarrow \infty$, $I$ does not approaches $N$ necessarily, depending on the recovery and death rates, $I$ does not reach $N$.

The last important quantity extracted from this model is the basic reproduction number $\mathcal{R}_0$, given by \cite{anastassopoulou2020data}:

\begin{align}
\label{eq.R0SIRD}
    \mathcal{R}_0 = \frac{\beta}{\gamma + \mu}.
\end{align}

This quantity, is of vital importance of the study of a disease outbreak.

\subsection{SEIRD}

Another deterministic mathematical model possible is the SEIRD model, in which we consider the population $N$ of a given region as divided in 5 groups. At time $t$, there are those who are susceptible to get infected $S(t)$, the ones who have already been exposed the virus but does not present symptoms yet $E(t)$, people who are already infected and present the symptoms $I(t)$, the ones that have already recovered from the disease $R(t)$ and those who are dead due to the infection $D(t)$. This model is a good approximation to a short epidemic, so the population of a region is roughly constant throughout the epidemic period. Also, since this is a deterministic model, we assume $N$ to be a big number compared to the number of people associated with the infection of a single person. The final consideration is that we also assume that people that are recovered from the disease acquire immensity and does not become susceptible to become infected again.

\begin{figure}[H]
    \centering
    \begin{tikzpicture}[node distance=3cm,
    every node/.style={fill=white, font=\sffamily}, align=center]
  % Specification of nodes (position, etc.)
  \node (start)             [activityStarts]              {Susceptible S(t)};
  \node (onCreateBlock)     [process, below of=start, xshift=2.5cm]          {Infected I(t)};
  \node (activityRuns)      [activityRuns, below of=onCreateBlock, xshift=-4cm]
                                                      {Recovered R(t)};
    \node (ActivityEnds)      [startstop, below of=onCreateBlock]
                                                        {Dead D(t)};
    \node (Exposed)         [process, below of=start, xshift=-2.5cm]      {Exposed E(t)};
                                    
  % Specification of lines between nodes specified above
  % with aditional nodes for description 
  \draw[->]             (start) -- node[text width=5cm]
                    {$\beta I(t)S(t) + kE(t)S(t)$} (Exposed);
  \draw[->]     (Exposed) -- node[text width=1cm]
                    {$cE(t)$} (onCreateBlock);
  \draw[->]     (onCreateBlock) --
  node[text width=1cm]
                    {$\gamma I(t)$}(activityRuns);
  \draw[->]     (onCreateBlock) -- node[text width=1cm]
                    {$\mu I(t)$}
  (ActivityEnds);
                                   
    \end{tikzpicture}
    \caption{Representation of a SEIRD model, a susceptible person gets exposed to the virus, being infected afterwards and either dies or recovers from the disease.}
    \label{fig.RepresentationSEIRD}
\end{figure}
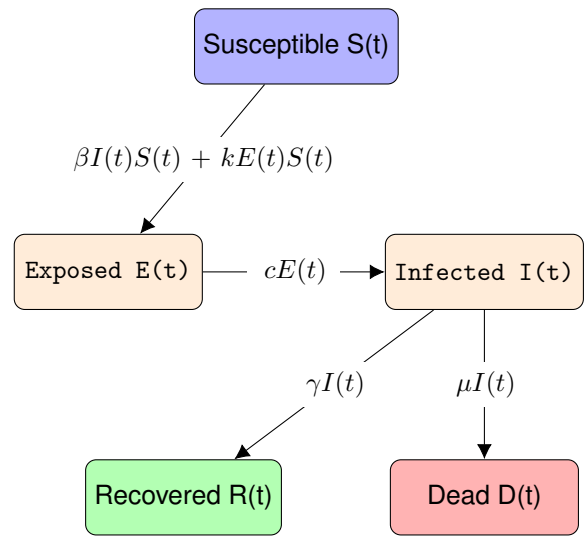

The rate of infection $\lambda$ is proportional to the number of people infected, $\lambda(t) = \beta I(t)$, where the constant $\beta$ represents the effectiveness of the infection, the rate of cure $\gamma = P_{:)}\tau_r^{-1}$, where $P_{:)}$ is the probability of recovery and $\tau_r$ is the average time taken for an infected person to recover. Similarly the rate of death is $\mu = P_{CFR} \tau_d^{-1}$, where $P_{CFR}=1-P_{:)}$ is the probability of death, given by the CFR and $\tau_d$ is the average time taken for an infected person to die. Figure \ref{fig.RepresentationSEIRD} carries an visual representation of the SEIRD model.

The differential equations representing the evolution of the populations are given by

\begin{align}
\label{eq.1}
    &\frac{dS}{dt} = -\frac{(1-P_{exp})\beta}{N} I(t) S(t) - \frac{P_{exp}\beta}{N}E(t)S(t)\\
\label{eq.2}
    &\frac{dE}{dt} = \frac{(1-P_{exp})\beta}{N} I(t)S(t) + \frac{P_{exp} \beta}{N}E(t)S(t) - cE(t)\\
\label{eq.3}
    &\frac{dI}{dt} = cE(t) - \gamma I(t) - \mu I(t) \\
\label{eq.4}
    &\frac{dR}{dt} = \gamma I(t) \\
\label{eq.5}
    &\frac{dD}{dt} = \mu I(t)
\end{align}

We first turn our attention to the construction of an appropriate formula for calculating $\mathcal{R}_0$ with this model. For that we follow the method derived on \cite{van2002reproduction}. The study develops a mathematical generalization for writing $\mathcal{R}_0$ depending on the type of epidemiological model. $\mathcal{R}_0$ is defined as

\begin{align}
    \mathcal{R}_0 = \rho(F V^{-1})
\end{align}

\noindent where $\rho(X)$ means the spectral radius of the matrix X, that is, the largest absolute eigenvalue. Both $F$ and $V$ are the matrices of the derivatives of the functions defining the behavior of the disease population, with respect to each population compartment.

To get to these matrices, we first note that the set of equations regarding the dynamics of the SEIRD model can be expressed as follow: Consider $\Vec{x}$ the vector of populations, that is $\Vec{x} = (x_1, x_2, x_3, x_4, x_5)$ where $x_1 = E$, $x_2 = I$, $x_3 = S$, $x_4 = R$ and $x_5 = D$. Analogously, $d\vec{x}/dt$ is the vector of the first derivatives. Then, we can write the dynamics of the populations as

\begin{align}
    \frac{d\vec{x}}{dt} = \mathcal{F} - \mathcal{V}
\end{align}

\noindent where $\mathcal{F}$ is the vector that relates the appearance of new infections on the disease populations due to contamination, and $\mathcal{V}$ is the input and output of members in all populations due to all other causes, such as recovery from the disease, development of symptoms after an incubation period, etc. In our case, since all newly infected members go to the $E$ population

\begin{align}
    \mathcal{F} = \begin{pmatrix}
    (1-P_{exp})\beta I S + P_{exp} \beta E S \\
    0 \\
    0 \\
    0 \\
    0
    \end{pmatrix}
\end{align}

\noindent while

\begin{align}
    \mathcal{V} = \begin{pmatrix}
    \beta cE \\
    -cE + \gamma I + \mu I \\
    (1-P_{exp})\beta I S +  P_{exp} \beta E S \\
    -\gamma I \\
    -\mu I
    \end{pmatrix}.
\end{align}

Now, we know that the situation of a disease free equilibrium (DFE), meaning no disease is happening, is achived by the vector $\vec{x}_0 = (0, 0, S_0, 0, 0)$, where $S_0 = N$. According now to \cite{van2002reproduction} we can calculate $F$ and $V$ as

\begin{align}
    &F = \left.\left( \frac{\partial \mathcal{F}_i}{\partial x_j} \right )\right|_{x = x_0} \hspace{1cm} 1 \leq i\\
    &V = \left.\left( \frac{\partial \mathcal{V}_i}{\partial x_j} \right )\right|_{x = x_0} \hspace{1cm} j \leq m\\
\end{align}

\noindent being $x_j$ the vector components of $\vec{x}$ related to the populations with the disease, in our case $E$ and $I$, and $m$ is the number of populations related to infectious beings. Here $m = 2$. Performing the derivatives, we conclude

\begin{align}
    &F = \begin{pmatrix}
    P_{exp} \beta & (1-P_{exp})\beta \\
    0 & 0
    \end{pmatrix} \\
    &V = \begin{pmatrix}
    c & 0 \\
    -c & \gamma + \mu
    \end{pmatrix}
\end{align}

The next step is to find the inverse matrix of $V$, fortunately $V$ is a 2x2 matrix and the formula for it's inverse is straightforward

\begin{align}
    V^{-1} = \frac{1}{c(\gamma + \mu)}\begin{pmatrix}
    \gamma + \mu & 0 \\
    c & c
    \end{pmatrix},
\end{align}

\noindent and we proceed to the last step of combining $FV^{-1}$ in order to retrieve $\rho(FV^{-1})$ and find $\mathcal{R}_0$.

\begin{align}
    &FV^{-1} = \frac{1}{c(\gamma + \mu)} \times \\
    \nonumber
    &\times \begin{pmatrix}
    P_{exp} \beta(\gamma + \mu) + (1-P_{exp})\beta c & (1-P_{exp})\beta c \\
    0 & 0
    \end{pmatrix},
\end{align}

\noindent therefore, by computing the eingenvalues of $FV^{-1}$ we find

\begin{align}
\label{eq.R0SEIRD}
    \mathcal{R}_0 = \frac{P_{exp} \beta\left[\gamma + (1-P_{exp})\beta\right] + (1-P_{exp})\beta c}{c(\gamma + \mu)}
\end{align}

Having $\mathcal{R}_0$ in our hands, we continue to the study of some behaviors of this model.

The set of equations describing the model is subjected to the initial conditions. When $t \rightarrow 0$, $I(t) - \rightarrow I_0$, $S(t) \rightarrow S = N - I_0$, $R(t) \rightarrow 0$, $D(t) \rightarrow 0$ and $E(t) \rightarrow E_0$, where $I_0$ is the initial number of infected, $E_0$ is the initial number of exposed in the population and no deaths or recoveries are assumed at $t = 0$.

\subsection{Non-pharmaceutical intervention}

\begin{figure}[H]
    \centering
    \includegraphics[width = 0.8\textwidth]{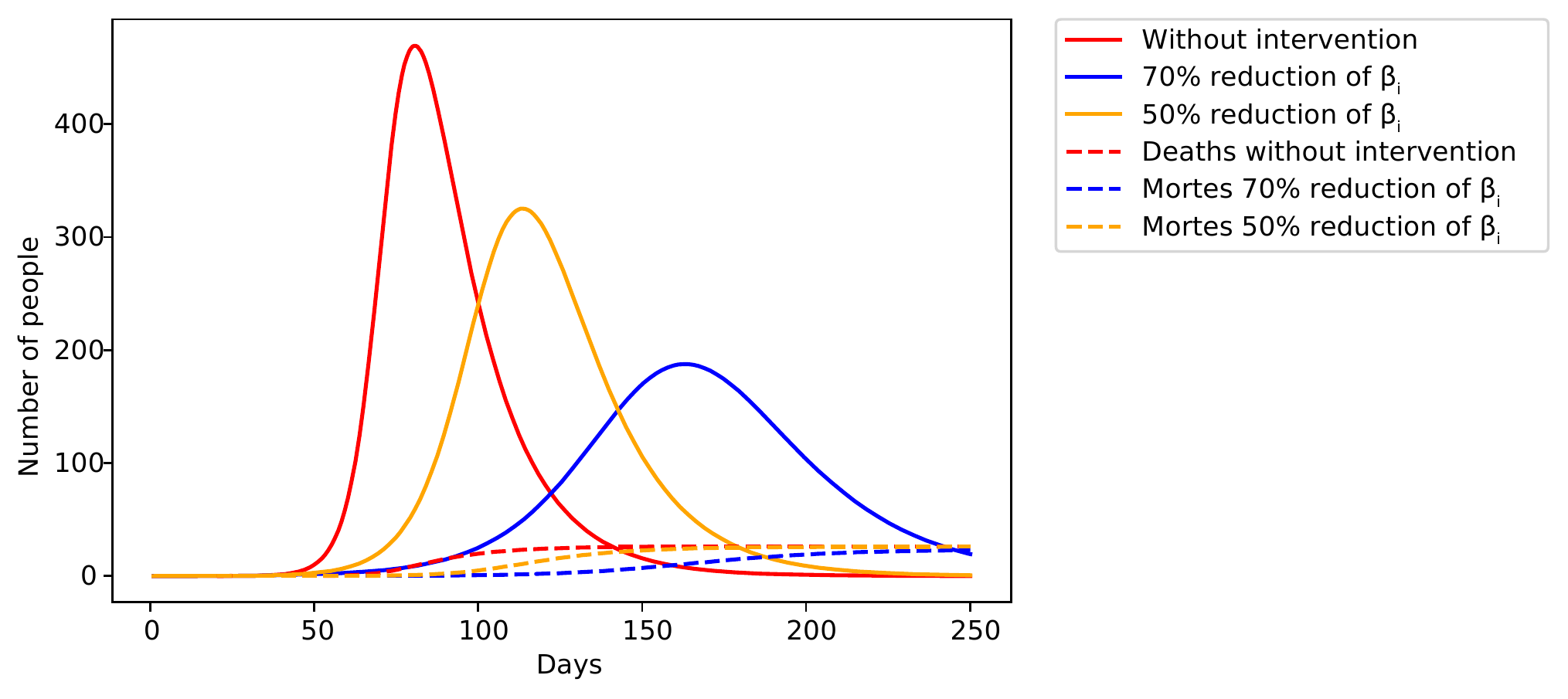}
    \caption{Visual representation of the effect of non-pharmaceutical interventions on the infection curve, depending on the efficiency of the intervention, given by $P_{dec}$.}
    \label{fig.intervention}
\end{figure}

Without vaccines or efficient medicine against the disease, non-pharmaceutical interventions are the only effective way to prevent further increase of the pandemic \cite{dehning2020inferring}. These interventions take different approaches such as social distancing, social isolation and lockdown of the population. Despite the differences, they all carry the same objective, decreasing the infection rate $\beta$. It is convenient to implement the effect of these interventions on the model, when making predictions. Here, we model this effect by a logistic function, where $\beta$ starts at a initial value $\beta_i$ and at some critical time $t_c$ a intervention is imposed and beta decreases to $\beta_f = P_{dec}\beta_i$, where $P_{dec}$ is the fraction of $\beta_i$ decreased by the intervention. In France, studies estimate that the intervention decreased $\beta_i$ by 77\% \cite{salje2020estimating}, therefore $P_{dec}= 0.77$ in France.

\begin{align}
    \beta(t) = \frac{(1-P_{dec})\beta_i}{1+\tau e^{t-t_c}} + P_{dec}\beta_i
\end{align}

where $\tau$ is a constant related to the time taken for the intervention to have the effect desired. Such model reconstruct the general behavior of interventions against the spread of the disease (Figure \ref{fig.intervention})

\subsection{Age division}

Since the case fatality rate (CFR) of COVID-19 is different among age groups \cite{surveillances2020epidemiological, wu2020characteristics, tian2020characteristics}, we propose here a modification on both models, including the age distribution of the population and the social aspects of close contact between members of the population. The modification is describe as follow: Each compartment is divided into $M$ age groups, where each $i$-th group has a $P_{CFR_i}$ associated to it, that is, the probability of death associated to the $i$-th age group. The $\beta$ parameter is now described as the average number of daily contacts between a member belonging to the $i$-th age group to the $j$-th age group, multiplied by the infection probability $P_{infc}$

\begin{align}
    \beta_{i_I} = \sum_{j=1}^N C_{ij} I_j P_{infc}, \\
    \beta_{i_E} = \sum_{j=1}^N C_{ij} E_j P_{infc},
\end{align}

\noindent where $C_{ij}$ is called the social contact matrix and we included $I_j$ and $E_j$ inside $\beta$ now to place everything on the same sum. The age distribution among the population is retrieved from the UN prospects \cite{UNprospect2019} and the social contact matrix for those countries was measured on previous studies \cite{mossong2008social, leung2017social}. The specific contact matrix for the Republic of Korea was not found, however, \cite{gupta2002cultural} finds evidences of cultural clusters in the world, where countries belonging to the same cluster share cultural similarities; thus, we use this fact to justify the use of Hong Kong's social contact matrix to describe the Republic of Korea. That way, we include cultural and population aspects for each of those countries, increasing the odds of a successful prediction. This type of model was used recently to describe the coronavirus outbreak on large cities in Brazil \cite{rocha2020expected}.

\section{Comparing Adjustments}

To test the SIRD and SEIRD model we first compare them to the pandemic crisis on the Republic of Korea, running a numerical solution for the differential equations \eqref{eq.1} - \eqref{eq.5} we adjust the general behavior of the populations to Korean data acquired from \cite{Worldeters} since 15/02/2020. The data from the Republic of Korea consists of the infection curve and death curve. To prevent problems with initial guess on the fitting process, both models used the same values for the initial guess, except $E_0$, which is found only on the SEIRD model. $S_0 = N$ was also left as a free parameter of the adjustment instead of set to the total population of the country, which is justified by a limitation in both models, where the population is assumed homogeneously spread, which does not correspond to reality. Thus, $N$ does not represent the total population, instead it represents an effective population smaller than the total population, due to non-homogeneous distribution throughout the territory, the interpretation of $N$ as the disease evolves is discussed on the discussion session. The parameter was chosen to be $k=0.44\beta$, we considered a study which estimated that presyntomatic cases caused 44\% of infections \cite{he2020temporal}, while for $c$ we used an average of several clinical studies shown on table \ref{tab.incubation}.

\begin{table}[H]
    \centering
    \begin{tabular}{|c|c|c|}
    \hline
    incubation time & 95\% confidence & Reference \\
    \hline
    6.4 days & 5.6-7.7 & \cite{backer2020incubation} \\
    \hline
    5.2 days& 4.1-7 & \cite{li2020early} \\
    \hline
    5 days & -- & \cite{linton2020incubation} \\
    \hline
    4 days & -- & \cite{guan2020clinical} \\
    \hline
    5.1 days & 4.5 - 5.8 & \cite{lauer2020incubation} \\
    \hline
    \end{tabular}
    \caption{Incubation time of the disease according to other studies.}
    \label{tab.incubation}
\end{table}

\begin{figure}[H]
  \centering
  \begin{minipage}[b]{0.49\textwidth}
    \includegraphics[width=\textwidth]{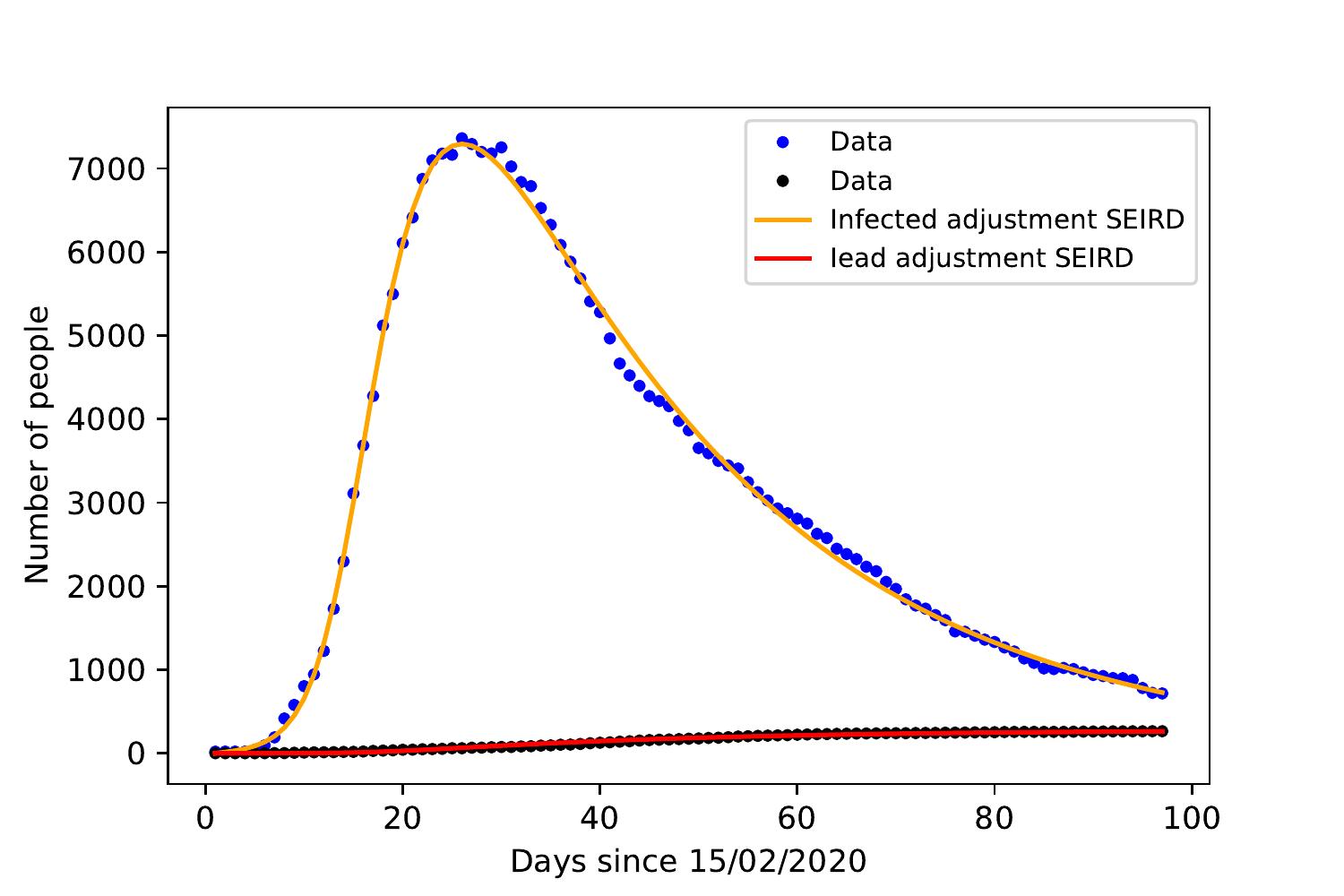}
    \caption{Fit for the infected and deaths by SARS-CoV-2 on the Republic of Korea using the SEIRD model.}
    \label{fig.CoreiaSEIRDfit}
  \end{minipage}
  \hfill
  \begin{minipage}[b]{0.49\textwidth}
    \includegraphics[width=\textwidth]{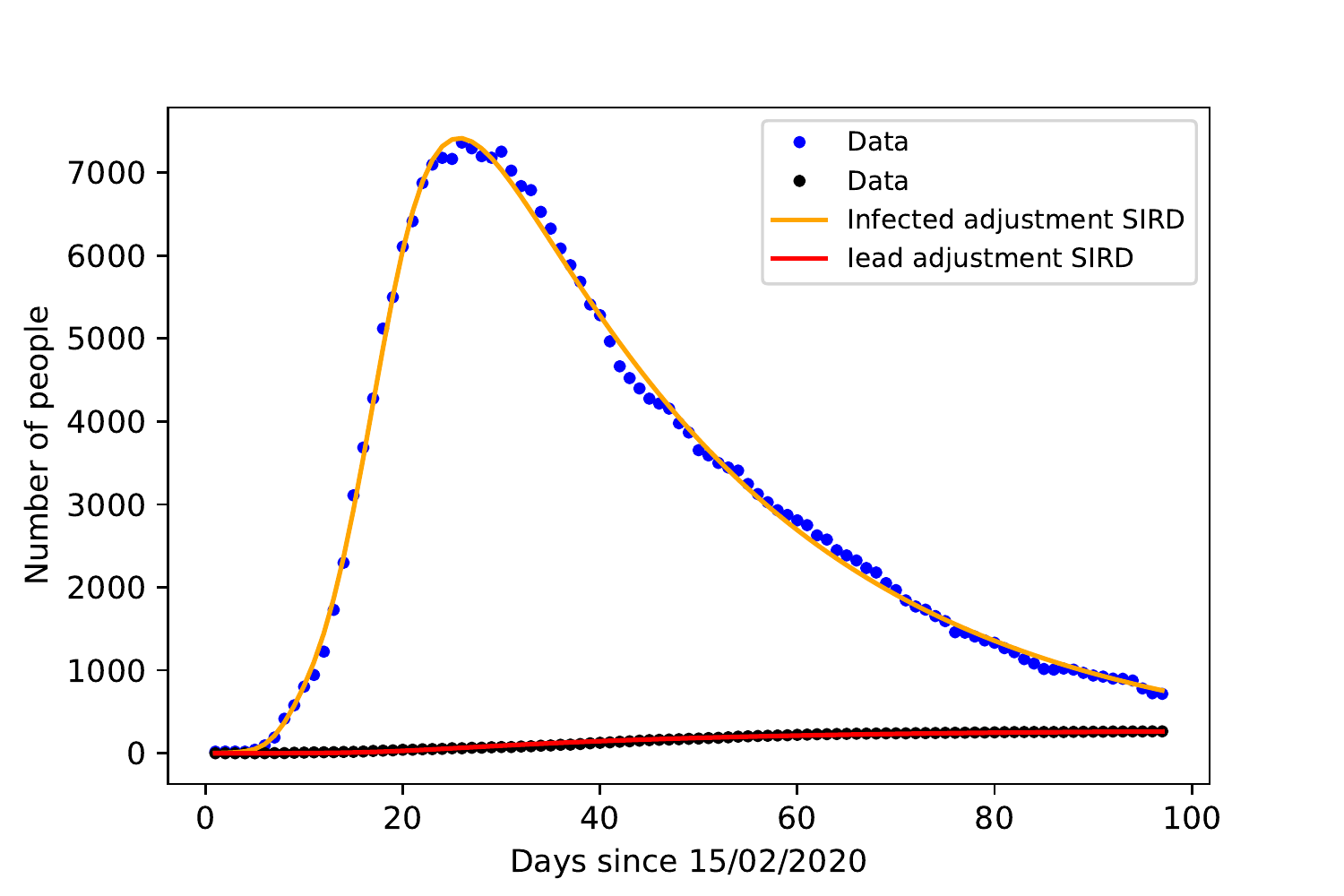}
    \caption{Fit for the infected and deaths by SARS-CoV-2 on the Republic of Korea using the SIRD model.}
    \label{fig.CoreiaSIRDfit}
  \end{minipage}
\end{figure}

Since the Korean government did not impose a lockdown or social isolation, we set $P_{dec} = 0$ in both models. Figures \ref{fig.CoreiaSEIRDfit} and \ref{fig.CoreiaSIRDfit} show the result of the fitting process and table \ref{tab.fitting} includes the acquired values for all parameters for each model.

\begin{table}[H]
    \centering
    \begin{tabular}{|c|c|c|}
    \hline
    Parameter & SIRD & SEIRD \\
    \hline
    $\chi^2$ & 0.9978 & 0.9978 \\
    \hline
    $\tau_r$ & 7.9 $\pm$ 0.3 & 8 $\pm$ 0.3 \\
    \hline
    $\tau_d$ & 29.6 $\pm$ 0.2 & 28.8 $\pm$ 0.2\\
    \hline
    $I_0$ & 2 $\pm$ 1 & 1 $\pm$ 4\\
    \hline
    $E_0$ & -- & 62 $\pm$ 57 \\
    \hline
    $\beta$ & 0.478 $\pm$ 0.004 & 0.513 $\pm$ 0.008\\
    \hline
    $N$ & 11035 $\pm$ 57 & 11218 $\pm$ 48\\
    \hline
    \end{tabular}
    \caption{Parameters found by the adjustment with both models}
    \label{tab.fitting}
\end{table}

The recovery time on both models is close to 8 days, while other studies such as \cite{bernheim2020chest} found 10 days. The time from symptoms onset to death was in both models close to 30 days, being 1 day shorter with the SEIRD model, \cite{ruan2020clinical} and \cite{Palmieri2020Report} found $\tau_d = 18$ or $11$ days.

Comparing the accuracy of the fitting with the data, both models resulted the same value of $\chi^2$. The parameter $E_0$ presents a large margin of error, which is expected given the lack of real data concerning the exposed population.

Proceeding to the calculation of $\mathcal{R_0}$ for both models, using equations \eqref{eq.R0SEIRD} and \eqref{eq.R0SIRD} we found

\begin{align}
    \mathcal{R_0}_{SEIRD} = 1.92 \pm 0.07\\
    \mathcal{R_0}_{SIRD} = 2.98 \pm 0.09
\end{align}

The value for $\mathcal{R}_0$ according to other studies ranges from 2 to 3 \cite{zhao2020preliminary, liu2020reproductive, zhang2020estimation, read2020novel}, therefore, both models yield acceptable values for $\mathcal{R_0}$ being the one predicted by the SEIRD model lower.

\section{Prediction Accuracy with Age Division}

We now proceed to test the prediction accuracy of both models. We used the first third of the data for fitting both models and extracting parameters, after having the parameters, we compare the prediction for the next days with these parameters with the rest of the dataset.

\subsection{Germany}

For Germany, the fitting data corresponds to the cases and deaths until 17th March. However, until the peak is reached, both models find presents very large margin of error for $N$, to avoid this problem, we varied $N$ manually, from 0 to 10\% of the local population; which was taken from a united nation prospect for the year of 2020 \cite{UNprospect2019}, at steps of 0.05\%. At each step, we fit the initial data and reject the fitting if the $\chi^2$ value is lower than 0.995, this $\chi^2$ method for validating the goodness of a data adjustment had already been used for epidemiological models \cite{tang2020prediction}. We than plotted the maximum and minimum acceptable fits to generate the margin of prediction, comparing it with the complete dataset. We also decided to use $\tau_r$ and $\tau_d$ according to clinical studies when performing the prediction, instead of leave them as free parameters for the fitting, $I_0$ was also determined a priori according to the first registered number on 15/02/2020. The resulting free parameters for fitting the training set are $P_{infc}$ and $E_0$.

With the SEIRD model, we found the maximum and minimum values of $N$ to be $N_{min} = 0.25\%$ of the German population and $N_{max} = 0.40\%$ of the German population. The $P_{infc}$ parameter varied from 15.5\% to 16.2\%

\begin{figure}[H]
  \centering
  \begin{minipage}[b]{0.49\textwidth}
    \includegraphics[width=\textwidth]{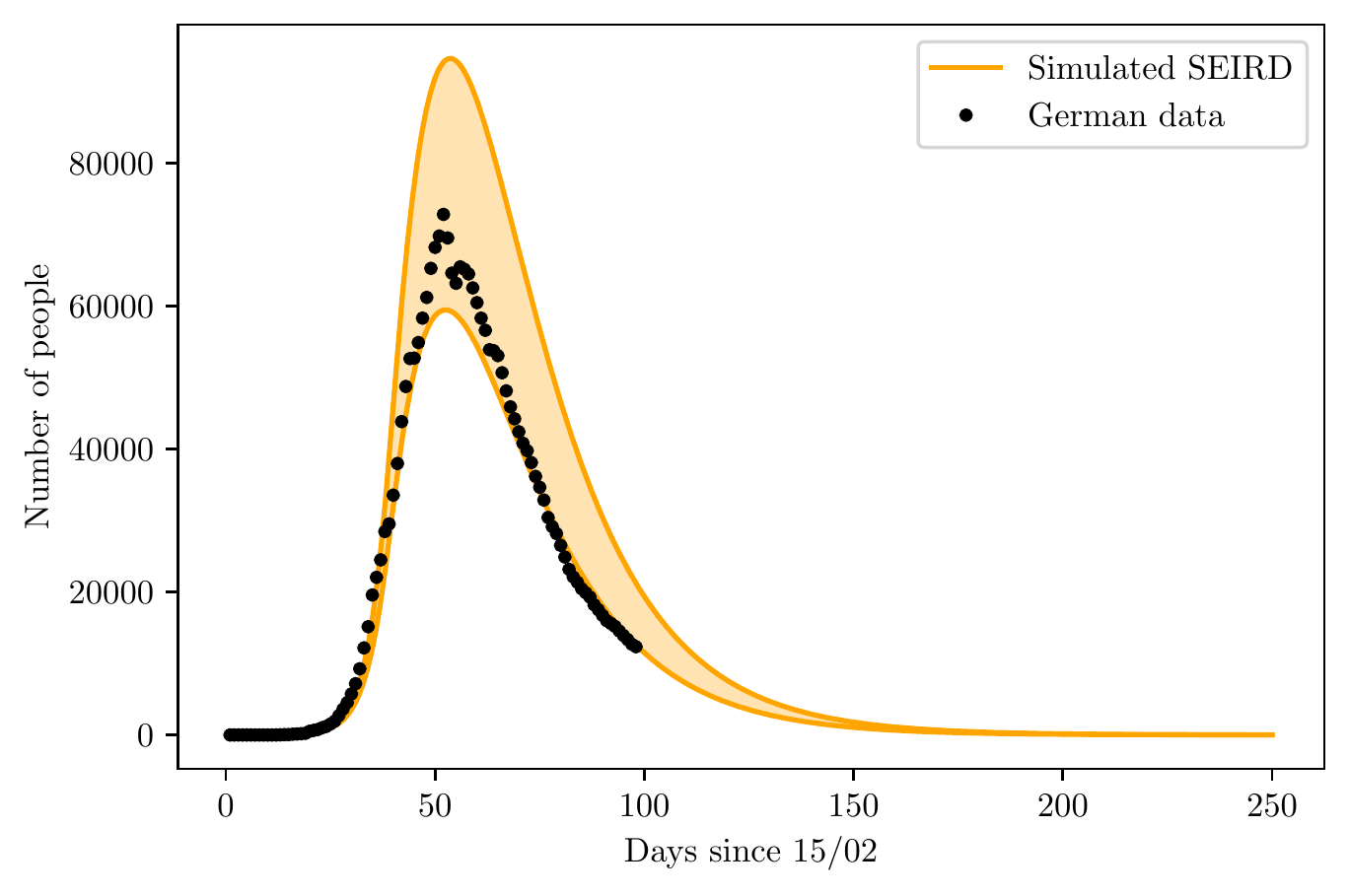}
    \caption{Prediction for Germany in comparison with real data using the SEIRD model with age division}
    \label{fig.AlemanhaSEIRD}
  \end{minipage}
  \hfill
  \begin{minipage}[b]{0.49\textwidth}
    \includegraphics[width=\textwidth]{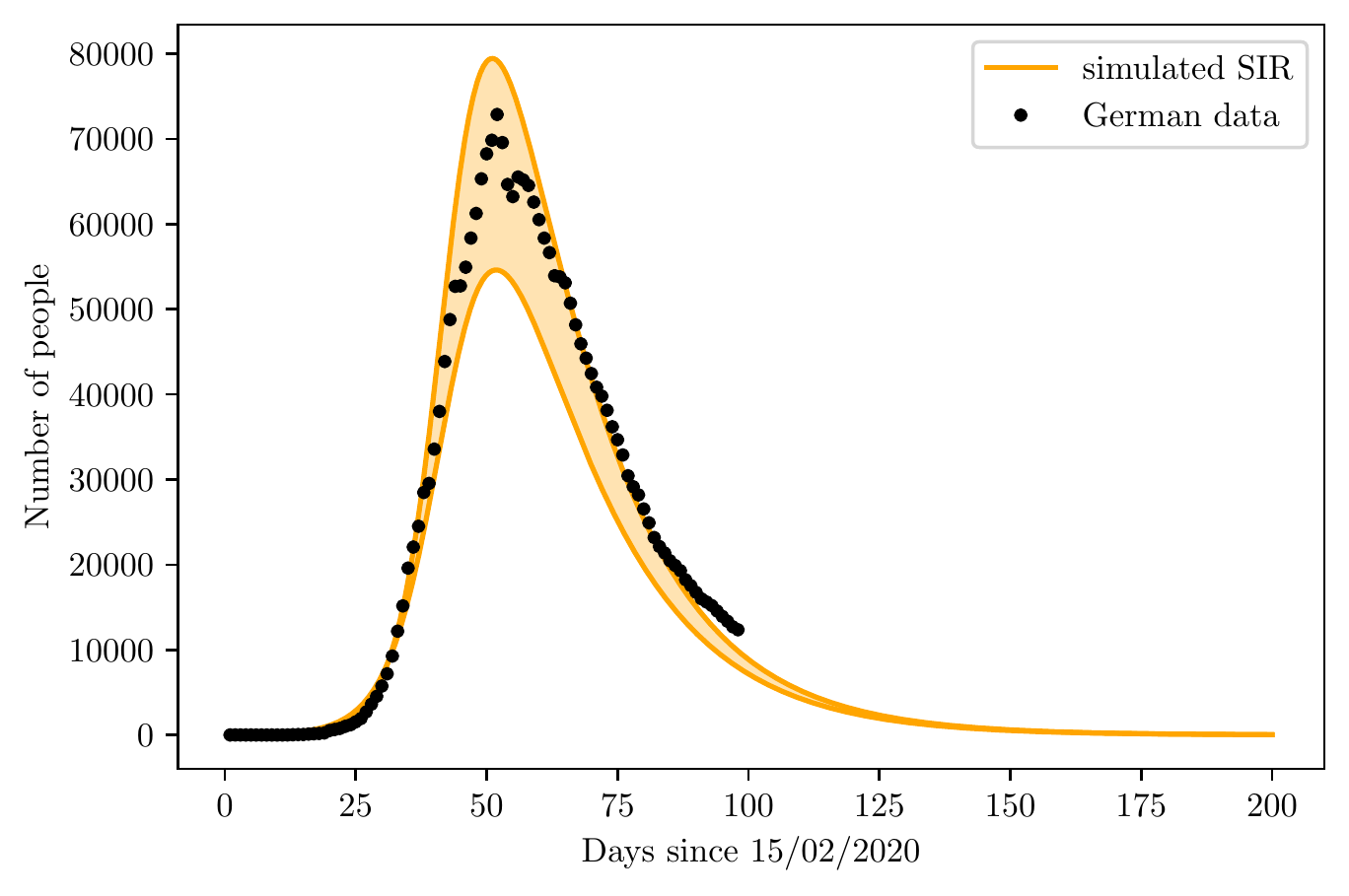}
    \caption{Prediction for Germany in comparison with real data using the simple SIRD model.}
    \label{fig.AlemanhaSIRD}
  \end{minipage}
\end{figure}

Using the SIRD, leaving $\beta$ and $I_0$ as free parameters. The limit values of $N$ were $N_{min} = 0.15\%$ and $N_{max} = 0.2\%$, while $\beta$ varied from $0.247$ to $0.279$ and $I_0$ from 11 to 30. Figures \ref{fig.AlemanhaSEIRD} and \ref{fig.AlemanhaSIRD} show the result of both simulations, with the maximum $N_{max}$ and minimum $N_{min}$ curves. The shaded region is the region between $N_{max}$ and $N_{min}$.

\subsection{Korea}

The training set consisted of 20 days, corresponding to the infections from 15/02 to 06/03. The SEIRD model found $N_{min} = 0.03\%$ and $N_{max} = 0.04\%$ of the total Korean population, while $P_{infc}$ varied between 80 to 85\%.

\begin{figure}[H]
  \centering
  \begin{minipage}[b]{0.49\textwidth}
    \includegraphics[width=\textwidth]{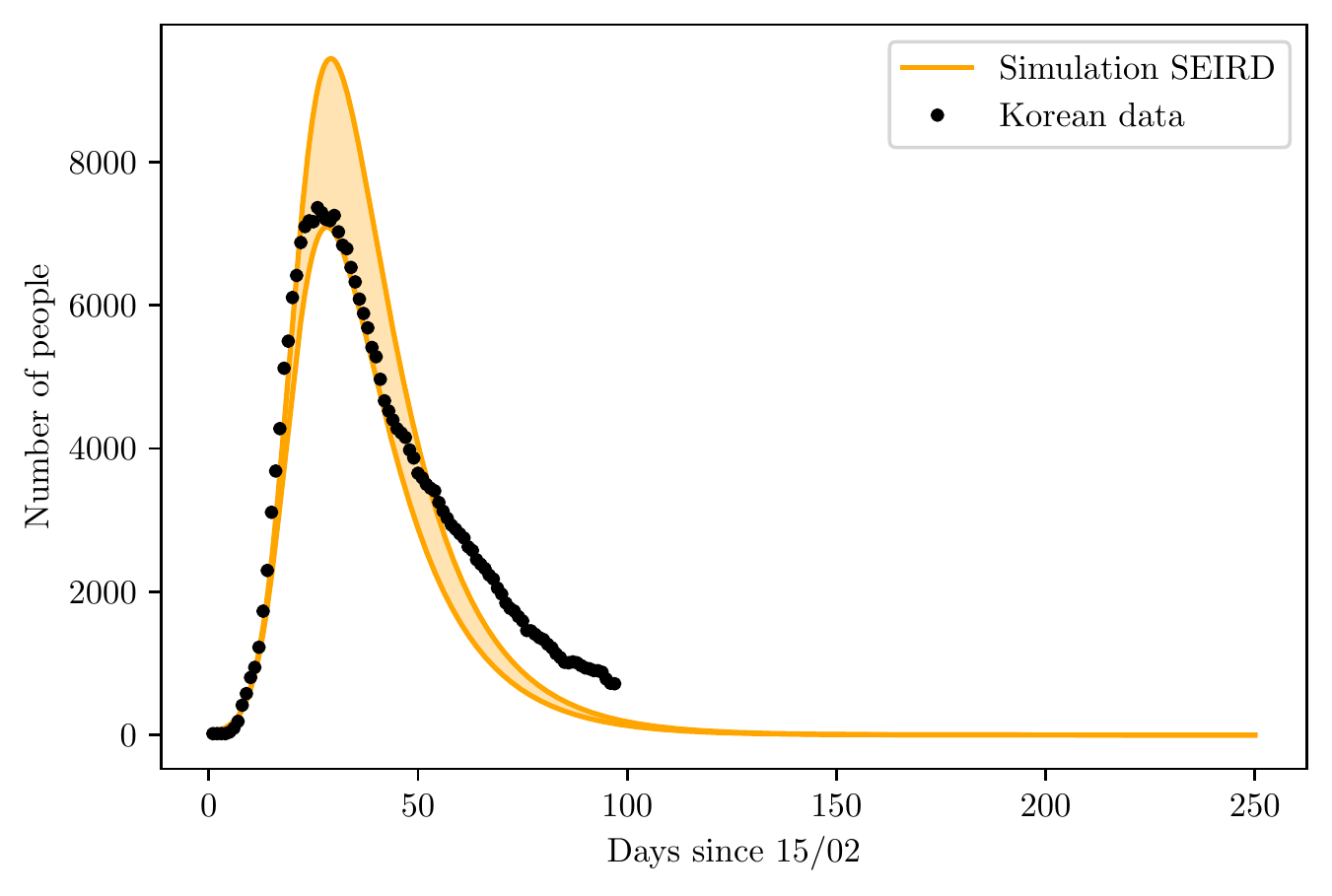}
    \caption{Prediction for the Republic of Korea in comparison with real data using the SEIRD model with age division}
    \label{fig.KoreaSEIRD}
  \end{minipage}
  \hfill
  \begin{minipage}[b]{0.49\textwidth}
    \includegraphics[width=\textwidth]{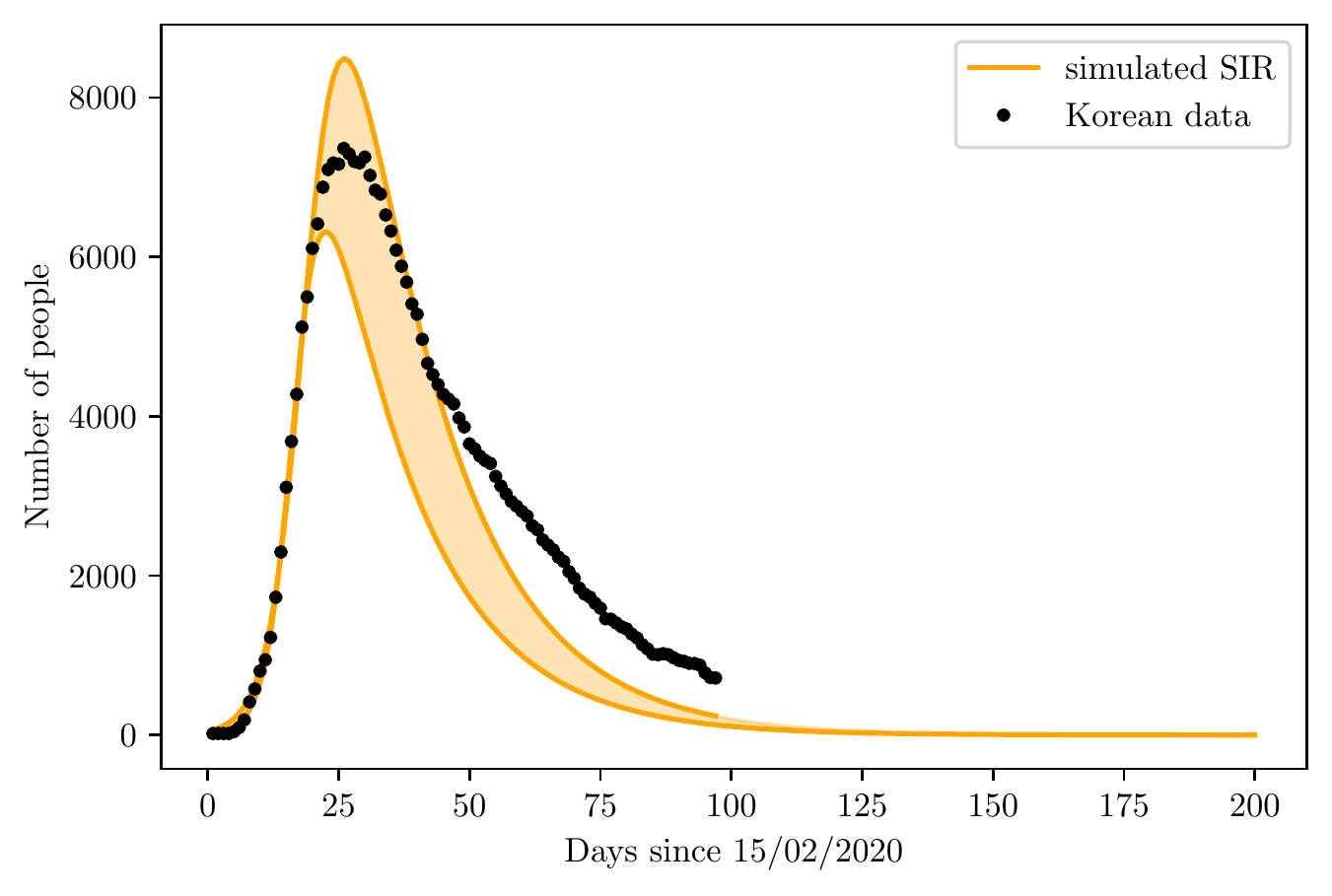}
    \caption{Prediction for the Republic of Korea in comparison with real data using the simple SIRD model.}
    \label{fig.KoreaSIRD}
  \end{minipage}
\end{figure}

The simple SIRD model found $N_{min} = 0.02\%$ and $N_{max} = 0.03\%$, $\beta$ went from 0.345 to 0.436, and $I_0$ was between 23 to 65. Figures \ref{fig.KoreaSEIRD} and \ref{fig.KoreaSIRD} present the result for prediction of both models.

\section{Discussion}

When concerning the adjustment process for acquisition of parameters with both models, there were no difference on the accuracy of the fit, and both models yielded very close values for the parameters. However, $\tau_d$ is super estimated in both models, being slightly lower on the SEIRD model. The value of $\tau_r$ is acceptable inside the variation of clinical measures.

\begin{figure}[H]
    \centering
    \includegraphics[width=0.9\textwidth]{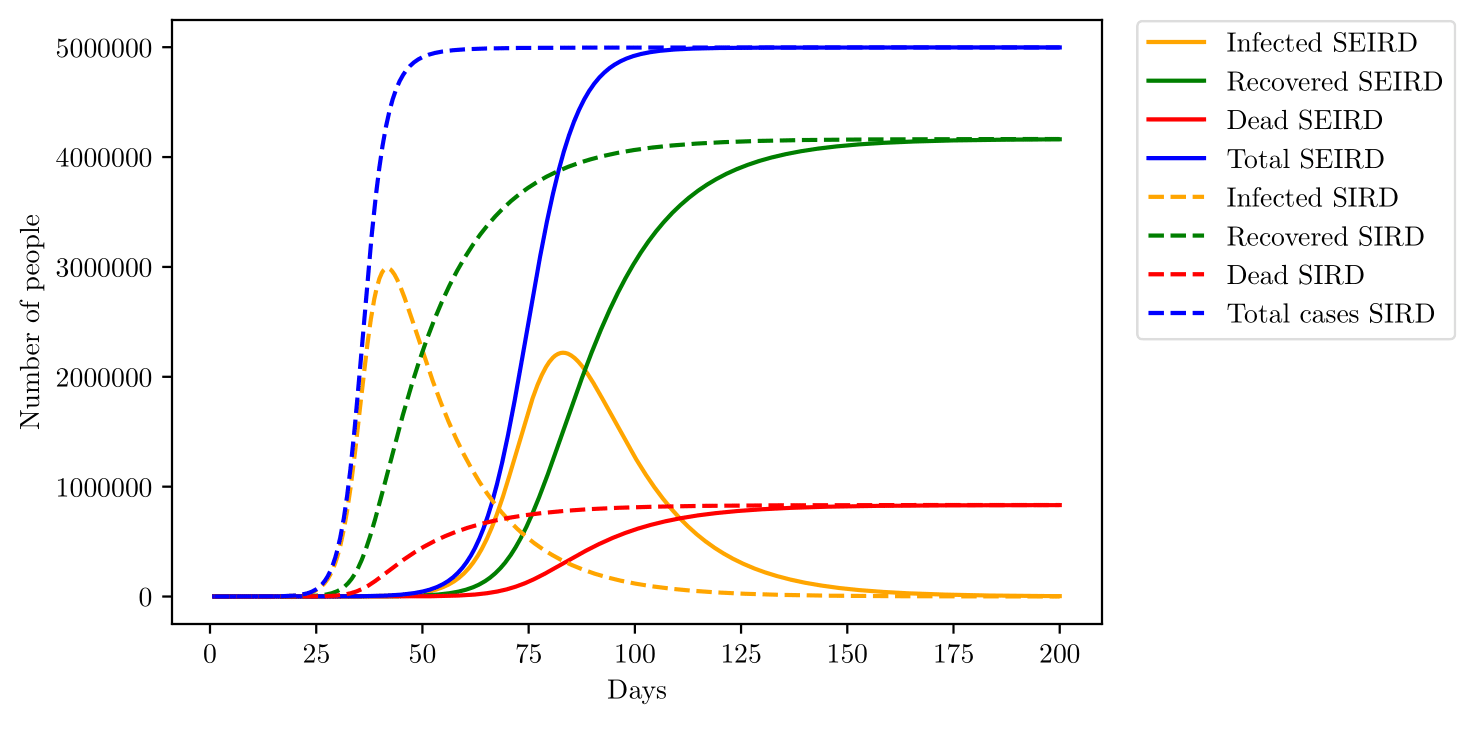}
    \caption{Comparison between SIRD and SEIRD models. The parameters chosen were the same for both models, except for $c = 1/5.1$ on the SEIRD. $\beta = 0.45$, $\gamma = 0.054$, $\mu = 0.0014$, $k = 0.44\beta$ and $N = 6000000$}
    \label{fig.comparison}
\end{figure}

The SEIRD model yields a slower growth rate than the SIRD model, that might happen due to the incubation period on the SEIRD model, which slows down the propagation of the virus towards other individuals. The main difference between the growth rate predicted by both models is better visualized by figure \ref{fig.comparison}, the action of the incubation period slows down the rate of infection, as seen by the adjustments, but also decreases the peak of infections. However, the cumulative numbers of infection, deaths and recoveries are the same.

$N$ could be understood as the population susceptible to the first pandemic wave, due to the non-homogeneous distribution of the population, not everyone is susceptible to the disease right at the start. With such an interpretation, $N$ tends to increase with time and approach the total population, here. Comparing predictions generated by the SIRD model with the SEIRD model with age division, the SEIRD model becomes a little more precise, although both simulations fail to predict the slower decrease of Korean data, that might be explained by an increase on $N$ as time passes, resulting in new cases registered and therefore, slowing down the rate of decrease. Such hypothesis is well acceptable since the Republic of Korea did not adopt any lockdown or social isolation measure, making the disease able to propagate towards other regions, increasing $N$ with time. Even with better prediction, the SEIRD model is far more complicated than the SIRD model and the use of the later should probably not compromise any data analysis. The same must hold true for simple SIR and SEIR models, when deaths are not a population to be accounted for, instead are just represented with a rate of removal for individuals.

The social isolation model developed here shows good results on the predictions, indicating that the description of $\beta$ should be close to reality. Here we find a huge advantage of the SEIRD model with age division in comparison with the SIRD model; by including age division, it is possible to simulate the effect of specific non-pharmaceutical interventions, such as school closure, which in principle would decrease $\beta_{i_I}$ and $\beta_{i_E}$ for the age groups between 0 to 19 years only. Another possibility is to include isolation of only elderly individuals. Several non-pharmaceutical measures have  been already described in literature \cite{lin2010optimal}, other studies show how the total number of infected might be changed due to the efficiency of non-pharmaceutical measures \cite{fergusonimpact}.

Other models might present more complete analysis of the disease, including hospitalizations and even asymptomatic cases, which are difficult to track and seem to vary a lot from place to place, the Diamond Princess cruise ship found 17.9\% of asymptomatic infections \cite{mizumoto2020estimating}, while an airplane flight found 11.2\% of cases being asymptomatic. An Italian village presented 50 to 70\% of cases being asymptomatic \cite{day2020covid}. There are yet the problem of assuring that the asymptomatic cases registered on studies are really asymptomatic and not presymtomatic, that is, are people still on the incubation period.

Of course, any mathematical model is only as good as the data allows, using mathematical models to describe the disease on countries with low testing rates might yield unrealistic predictions. For example, \cite{li2020substantial} estimates 86\% of infections being undocumented on China, at the early stages of the outbreak.

Another consideration we did not take, was the possibility of reinfection, where individuals leave the recovered group and re-enter the susceptible compartment. However, since other coronaviruses belonging to the same genus \textit{betacoronavirus} such as the SARS-CoV and the MERS-CoV does not present a high enough mutation rate to cause reinfection in short term \cite{zhao2004moderate}, the only cause of reinfection would be the loss of antibodies to fight the virus; nevertheless, on both diseases, the infected person acquires antibodies enough to prevent reinfection for a period of 2 - 3 years \cite{liu2006two}. With those considerations, we did not assume reinfection was probable on short-term. Future studies may be conducted to study the possibility of reinfection of individuals on the long-term.

\section{Conclusion}

Mathematical models of a disease outbreak such as the COVID-19 are able to predict the behavior of the infection. Both models have proved to be efficient tools for acquiring data and forecast the future situation. Despite the limitations, the models made it possible to achieve a value of $\mathcal{R}_0$ in good agreement with other studies, providing evidence in favor of the validity of the model.

However, the present models do not take into consideration the spatial distribution of the population, reflecting on some uncertainties that made the window of prediction larger.

The age division does not change the prediction drastically, suggesting that in the case of a simple prediction or analysis, SIRD models are useful. The age division SEIRD model provides an advantage when requiring specific simulations on specific groups of the population.

\bibliographystyle{unsrt}  
\bibliography{references.bib}  %%% Remove comment to use the external .bib file (using bibtex).
%%% and comment out the ``thebibliography'' section.

%%% Comment out this section when you \bibliography{references} is enabled.

\end{document}